\documentstyle[12pt]{article}
\renewcommand{\thesection}{\arabic{section}}
\renewcommand{\theequation}{\thesection.\arabic{equation}}
\title{(2+1)-Gravity and the embedding its dynamiycal symmetry and
para-supersymmetry into $SO(4,c)$ group}
\author{M.A. Jafarizadeh$^{a,b,}$\thanks{Email:Jafarzadeh@ark.tabrizu.ac.ir} ,
H. Fakhri$^{a,}$\thanks{Email:Hfakhri@ark.tabrizu.ac.ir}  and
S.K. Moayedi$^{a,}$\thanks{Email:Moayedi@ark.tabrizu.ac.ir}   \\
$^a${\small Faculty of Physics, Tabriz University, Tabriz, 51664, Iran.} \\
$^b${\small Institute for Studies in Theoretical Physics and Mathematics,
Tehran, 19395-1795, Iran.}}
\begin{document} 

\maketitle
\begin{abstract}

Some special solutions of the Einstein-Maxwell action with a non-negative 
cosmological constant and a very heavy point mass particle have been obtained.
The solutions correspond to static spacetime of locally constant curvature 
in its spatial part and a constant magnetic field of a magnetic monopole 
together with deficit of angle at the location of point mass. The quantum mechanics of a 
point particle in these spacetimes in the absence of angular deficit has
been solved algebraically both
relativistically and  non-relativistically. It has been also shown that these 
2-dimensional Hamiltonians have the degeneracy group of $GL(2,c)$ type and 
para-supersymmetry of arbitrary order or shape invariance, which is originated 
from a $SO(4,c)$ group.

\vspace{0.2cm}


{\bf PACs Index: 02.30. +g  or  02.90. +p.}

\end{abstract}

\vspace{0.5cm}

\begin{center}
\section{INTRODUCTION}
\end{center}

Due to vanishing of Weyl tensor in $(2+1)$ dimensions, the  Riemann curvature    
tensor can be identified with zero in matter-void regions    
of spacetime. Consequently, spacetime is flat in local vacua.
Addition of cosmological constant term in the absence of matter leads to
solutions with constant curvature, where the sign of cosmological constant 
determines the sign of scalar curvature \cite{Jackiw}.
In the last few years many interesting problems have been investigated in $(2+1)$
dimensional gravity, such as the spacetime metric of multi-point particle with
or without spin \cite{Deser,Clement}. The Einstein-Maxwell equations in $(2+1)$ dimensions
have already been treated \cite{Kogan,Reznik}. 
Indeed, in many physical situations
in $(3+1)$ dimensions there is no structure along one of spatial dimension like
an infinite cosmic string, where the theory becomes $(2+1)$ dimensional.
There are also some interesting works concerning the quantum mechanics of
a point mass particle in the presence of a very heavy particle in $(2+1)$
dimensions, both relativistically and nonrelativistically \cite{Dese,Sousa,Krzysztof}.

We introduce the Einstein-Maxwell action in the presence of
matter together with cosmological term in $(2+1)$ spacetime dimension. 
Then, we choose the solutions that correspond to a spacetime with a spatial part of 
locally constant curvature surface with deficit of angle at location of a very heavy
point mass, and magnetic field of a magnetic monopole.  
For positive cosmological
term we have Minkowskian and Euclidean spacetimes with spatial part of 
locally constant curvature. For vanishing cosmological term
we have a spacetime with locally flat spatial part.
Over these spacetimes, in absence of angular deficit,  the quantum Hamiltonian
associated with a free test particle leads to solvable systems with
degeneracy group $GL(2,c)$, where their eigen-states can be obtained
algebraically, too.
We will show that quantum solvable models will be obtained
by restricting the Casimir of $SO(4,c)$ group to the Casimir of $SL(2,c)$ group.
These models possess simultaneously the degeneracy group $SL(2,c)$ and shape
invariance symmetry, where both symmetries are kind of realization of
para-supersymmetry of arbitrary order.

\begin{center}
\section{SOME SPECIAL EXACT SOLUTIONS OF EINSTEIN EQUATIONS}
\end{center}

In $(2+1)$ dimensions the Einstein-Hilbert action of gravity coupled to matter 
and electromagnetic field, together with the cosmological term can be written as   
\begin{equation}
S=\int d^3x \sqrt{-g} \{ \frac{1}{4 \pi G}(R+2\Lambda) + 
\frac{1}{4} F_{\mu\nu}F^{\mu\nu}-{\cal L}_M \},
\end{equation}
where $\cal L_{M}$ is the matter Lagrangian. We have rescaled G by a 
factor of $4$.  
Variation of the action (2.1) leads to Einstein-Maxwell equations in
$(2+1)$ dimensions 
\renewcommand{\theequation}{\thesection.\arabic{equation}{a}}
\begin{equation}
R_{\mu\nu}-\frac{1}{2}g_{\mu\nu}R=2\pi GT_{\mu\nu}^{eff} 
\end{equation}
\vspace{-7mm}
\renewcommand{\theequation}{\thesection.\arabic{equation}{b}}
\setcounter{equation}{1}
\begin{equation}
\partial_\mu(\sqrt{-g}F^{\mu\nu})=0
\end{equation}
with
$T_{\mu\nu}^{eff}=T_{\mu\nu}^{(M)}+T_{\mu\nu}^{(EM)}+\frac{\Lambda}{2\pi G} g_{\mu\nu}$.
The energy-momentum tensor of matter $T_{\mu\nu}^{(M)}$ and electromagnetic 
$T_{\mu \nu}^{(EM)}$ are respectively
\renewcommand{\theequation}{\thesection.\arabic{equation}{a}}
\begin{equation}
T_{\mu\nu}^{(M)}=\frac{2}{\sqrt{-g}}\frac{\delta (\sqrt{-g}{\cal L}_{M})}{\delta g^{\mu\nu}},
\end{equation}
\vspace{-7mm}
\renewcommand{\theequation}{\thesection.\arabic{equation}{b}}
\setcounter{equation}{2}
\begin{equation}
T_{\mu\nu}^{(EM)}=-g^{\alpha \beta}F_{\mu \alpha} F_{\nu \beta}+\frac{1}{4}
g_{\mu\nu}F_{\lambda \sigma} F^{\lambda \sigma}.
\end{equation}

We assume that our spacetime is described by the axial symmetric static metric, that is \hspace{10mm}
$(\partial_{t}g_{\mu\nu}=0, g_{_{0i}}=0)$  \cite{Jack}: 
\renewcommand{\theequation}{\thesection.\arabic{equation}}
\begin{equation}
ds^2_{(3)}=N^2({\bf r})dt^2-\rho({\bf r})(dr^2+r^2d\phi^2),
\end{equation}
where $0\leq r<\infty$ and $0\leq\phi<2\pi$ are the usual polar coordinates.
The non-zero components of electromagnetic field tensor are 
$$
F_{0r}=E(r),\hspace{6mm} F_{r\phi}=B(r) \hspace{6mm} and \hspace{6mm} F_{0\phi}=0
$$ 
with E and B as electric and magnetic fields, respectively.
After, changing the coordinates as
$x^1=r \cos \phi $ and $x^2=r \sin \phi $, 
the electromagnetic energy-momentum tensor (2.3b)  
takes the following form
\begin{equation}
T_{\alpha \beta }^{(EM)}=\left(\begin{array}{ccc}
\frac{1}{2 \rho}(\frac{N^2}{\rho r^2}B^2+E^2)  &  \frac{x^2}
{\rho r^2}EB &  \frac{-x^1}{\rho r^2}EB  \\
\frac{x^2}{\rho r^2}EB  &  \frac{1}{2\rho r^2}B^2+\frac{(x^2)^2-(x^1)
^2}{2 r^2 N^2} E^2 & \frac{-x^1 x^2}{N^2r^2}E^2  \\ 
\frac{-x^1}{\rho r^2}EB & \frac{-x^1x^2}{N^2r^2}E^2 & \frac{1}{2\rho r^2}B^2-
\frac{(x^2)^2-(x^1)^2}{2r^2N^2}E^2  \end{array}\right). 
\end{equation}
In this article we consider the matter as a point mass located at the 
origin $(x^1=x^2=0)$ with the only nonvanishing component of $T_{\mu \nu}^{(M)}$ as
$T_0^{(M) 0}=\frac{1}{\sqrt{^{(2)}g}} M \delta (x^1) \delta (x^2)$, 
where $^{(2)}g$ is the determinant of  the  spatial  part  of  the  metric.  
We remind that the Ricci scalar for metric (2.4) is
\begin{equation}
R=\frac{2}{\rho N}\nabla^{2}N+\frac{1}{\rho^{2}}\nabla^{2} \rho - 
\frac{1}{\rho^{3}}(\vec{\nabla} \rho)^{2},
\end{equation}
where $\nabla^{2}$ and $\vec{\nabla}$ are the usual Euclidean 2-dimensional
Laplacian and gradient operators, respectively. 
For a spatially conformal metric with  
$\rho({\bf r})=\rho_{_{_{0}}}e^{-2GM \ln r+\chi}$,
the singular term on the right hand side of Eqs. (2.2a) disappears.
Thus, the equations can be written in the following singularity free form
\begin{eqnarray}
&& EB=0 \nonumber  \\
\nonumber  \\
&& \nabla^{2} \chi +2 \Lambda \rho_{_{_{0}}}r^{-2GM} e^{\chi}+\frac{2\pi G}{\rho_{_{_{0}}}}r^{2GM-2} 
e^{-\chi}B^{2}+\frac{2\pi G}{N^{2}}E^{2}=0  \nonumber  \\
\nonumber   \\
&& \partial_{1}\partial_{2}N-\frac{1}{2}(\partial_{1}N\partial_{2}\chi 
+\partial_{1}\chi \partial_{2}N)+\frac{GM}{r^{2}}(x^{2}\partial_{1}N
+x^{1}\partial_{2}N)-2\pi G\frac{x^{1}x^{2}}{Nr^{2}}E^{2}=0  \nonumber   \\
\nonumber   \\
&& \partial_{2}^{2}N+\Lambda \rho_{_{_{0}}}Nr^{-2GM} e^{\chi}+\frac{1}{2}(
\partial_{1}\chi \partial_{1}N-\partial_{2} \chi \partial_{2}N)-GM
r^{-2}(x^{1}\partial_{1}N-x^{2}\partial_{2}N) \nonumber \\
&&\hspace{45mm} -\frac{\pi G}{\rho_{_{_{0}}}}Nr^{2GM-2}e^{-\chi}B^{2}-\pi Gr^{-2}\frac{(x^{2})^{2}-(x^{1})^{2}}{N}E^{2}=0   \nonumber   \\
\nonumber    \\
&&\partial_{1}^{2}N+\Lambda \rho_{_{_{0}}}Nr^{-2GM}e^{\chi}+\frac{1}{2}(
\partial_{2} \chi \partial_{2}N-\partial_{1} \chi \partial_{1}N)-GM 
r^{-2}(x^{2}\partial_{2}N-x^{1}\partial_{1}N) \nonumber \\
&&\hspace{45mm}-\frac{\pi G}{\rho_{_{_{0}}}}Nr^{2GM-2}e^{-\chi}B^{2}-\pi Gr^{-2}\frac{(x^{1})^{2}-(x^{2})^{2}}{N}E^{2}=0. 
\end{eqnarray}

Letting $E=0$ in Eqs. (2.7) and using the following
ansatz for the magnetic field $B$
$$
B^{2}=h \frac{\Lambda \rho_{_{_{0}}}^2}{\pi G} r^{2-4GM}e^{2\chi} 
$$                            
where $h$ is a constant parameter. Then, with a change of variable 
$u=r^{1-GM}$, the Eqs. (2.7) can be written as 
\renewcommand{\theequation}{\thesection.\arabic{equation}{a}}
\begin{equation}
\frac{1}{u}\frac{d}{du}(u\frac{d}{du})\chi + 
\frac{2(1+h)\Lambda \rho_{_{_{0}}}}{(1-GM)^{2}}e^{\chi}=0   
\end{equation}
\vspace{-3mm}
\renewcommand{\theequation}{\thesection.\arabic{equation}{b}}
\setcounter{equation}{7}
\begin{equation}
\frac{1}{u}\frac{dN}{du}=\frac{c}{1-GM}e^{\chi}  
\end{equation}
\vspace{-3mm}
\setcounter{equation}{7}
\renewcommand{\theequation}{\thesection.\arabic{equation}{c}}
\begin{equation}
\frac{1}{u}\frac{d}{du}(u\frac{d}{du})N+ 
\frac{2(1-h)\Lambda \rho_{_{_{0}}}}{(1-GM)^{2}}Ne^{\chi}=0,  
\end{equation}
where $c$ is a constant of integration. Choosing the following ansatz as a solution for
the Eq. (2.8a)  
\renewcommand{\theequation}{\thesection.\arabic{equation}}
\begin{equation}
e^{\chi}=\frac{1}{(1+\frac{(1+h)\Lambda \rho_{_{_{0}}}}{4(1-GM)^{2}}u^{2})^2} ,   
\end{equation}
we make use of (2.9) to solve the Eq. (2.8b) as 
$$
N=c\frac{\frac{-2(1-GM)}{(1+h)\Lambda \rho_{_{_{0}}}}}{1+\frac{(1+h)\Lambda \rho_{_{_{0}}}}{4(1-GM)^{2}}u^{2}}+d,    
$$
where $d$ is another constant of integration.
Next, having inserted the results just obtained for $\chi$ and $N$ in
Eq. (2.8c), we get the following equations for constants  $d$ and $c$
\begin{eqnarray}
&&\frac{(1-h^{2})\Lambda \rho_{_{_{0}}}}{1-GM}d-(1+h)c=0  \nonumber \\
&&\frac{(1-h)\Lambda \rho_{_{_{0}}}}{1-GM}d-\frac{1-3h}{1+h}c=0.
\end{eqnarray}
The nontrivial solutions of Eqs. (2.10) exist only for  $h=0$ and $1$.                                               
If $h=0$, the magnetic field vanishes and Maxwell Eqs. (2.2b)
are satisfied. Then, (2.4)
becomes the metric of spacetime in the presence of a point mass located at the origin,
together with the cosmological term which has already been studied 
in Refs. \cite{Jackiw,Reznik}.

But, for $h=1$ we have $N=1$ and
\begin{equation}
B^{2}=\frac{\Lambda \rho_{_{_{0}}}^2}{\pi G} \frac{r^{2-4GM}}{(1+\frac{\Lambda \rho_{_{_{0}}}}{ 
2(1-GM)^{2}}r^{2-2GM})^{4}},
\end{equation}
together with the following metric of spacetime 
\begin{equation}
ds^{2}_{(3)}=dt^{2}-\frac{\rho_{_{_{0}}}r^{-2GM}}{(1+\frac{\Lambda \rho_{_{_{0}}}}{2(1-GM)^2}
r^{2-2GM})^2}(dr^{2}+r^2d\phi^{2}).
\end{equation}
It turns out that the magnetic field (2.11) and the metric (2.12) satisfy
Einstein-Maxwell Eqs. (2.2).     
Using the formula (2.6) we calculate the scalar curvature of metric (2.12)
$$
R=-4\Lambda-4\pi GMr^{2GM}(1+\frac{\Lambda \rho_{_{_{0}}}}{2(1-GM)^2}r^{2-2GM})^2\delta^2(\bf r).
$$
Hence, except for a delta singularity at the origin, the spacetime has a 
constant curvature. We will discuss the
solutions of Einstein equations in $(2+1)$-dimensional spacetime
corresponding to a point mass located at origin in the presence of cosmological
constant and magnetic field in the next section.
In the zero limit of cosmological constant, the magnetic field
(2.11) vanishes and the metric (2.12) changes to
the metric of a point mass located at origin
with angular deficit \cite{Deser,Jac}.
Note that, we have considered here only the most simple
solution of Liouville equation (2.8a), while one can take some less trivial solutions
using the Backlund transformations.

\begin{center}
\section{EMBEDDING OF DEGENERACY AND SHAPE INVARIANCE OF QUANTUM STATES
IN $SO(4,c)$ GROUP}
\end{center}
\setcounter{equation}{0}

We investigate the solutions (2.12) of Einstein equation 
in $(2+1)$ dimensions with nonnegative cosmological constant $\Lambda$, point mass $M$ 
and magnetic field (2.11) as sources of energy-momentum tensor. 
Now, we introduce
the parameter  $\gamma$  which only takes the values $0$, $1$ and
$i=\sqrt{-1}$ and redefine the cosmological constant $\Lambda$ as 
$\Lambda=\gamma^2 \lambda$.
In this article, $\lambda$ and $\rho_{_{_{0}}}$ are arbitrary positive (negative) nonvanishing constants
for $\gamma=0$ and $1$ ($i$). With change of variables 
\begin{eqnarray}
&&\frac{r^{1-GM}}{1-GM}=\sqrt{\frac{2}{\lambda \rho_{_{_{0}}}}} \frac{\tan \frac{\gamma \theta}{2}}{\gamma} \nonumber \\
&&\Phi:=(1-GM) \phi,
\end{eqnarray}
the metric (2.12) takes the form
\begin{equation}
ds^{2}_{(3)}=dt^{2}-\frac{1}{2 \lambda}(d \theta^{2}+\frac{\sin^{2}\gamma \theta}
{\gamma^{2}}d\Phi^{2}).
\end{equation}
It is obvious that for $\gamma=i$ the spacetime is described  
by an Euclidean metric, while for $\gamma=0$ and $1$ the spacetime is described
by Minkowskian metrics.
Using the general coordinate transformation, the magnetic field can be written 
as (3.3) in terms of the new coordinates $\theta$ and $\Phi$ 
\begin{equation}
{\cal B}=q\gamma \sin \gamma \theta,
\end{equation}
where
\begin{eqnarray}
\nonumber
q=\left \{\begin{array}{ll}
\frac{1}{2\sqrt{\pi G\lambda}}   & \mbox{if $\gamma=0$ , $1$}      \\
\frac{-1}{2\sqrt{\pi G|\lambda|}} & \mbox{if $\gamma=i.$}\end{array} \right.
\end{eqnarray}
The magnetic potential one-form $A$ corresponding to magnetic field (3.3) is
$A=q(1-\cos \gamma \theta)(\frac{i\gamma}{\sin \gamma \theta}d\theta +d\Phi)$. 
Obviously, in terms of the new coordinates $\theta$ and $\Phi$, the choice of $\gamma
=0$ leads to a Minkowskian metric with flat spatial part and 
angular deficit  in its metric. Here, $\theta$ is its 
radial coordinate and the magnetic field $\cal B$ is zero.
The choice of $\gamma=1$ leads to Minkowskian metric 
of $(2+1)$-dimensional spacetime with local constant curvature 
together with angular deficit and magnetic field $\frac{1}{2\sqrt{\pi G\lambda}} \sin \theta$.
Finally, for $\gamma=i$, we get Euclidean metric of 
$(2+1)$-dimensional spacetime with local constant curvature and 
deficit of angle and magnetic
field $\frac{1}{2\sqrt{\pi G|\lambda|}}\sinh \theta$.

In the presence of the magnetic field (3.3), the quantum states of
a point mass with mass ${\cal M}$, which is 
negligible compared to the mass $M$ of a point source located at the origin,
i.e. ${\cal M} \ll M$, 
can be described in terms of the energy spectrum of the 
Hamiltonian
$H=\frac{-1}{2{\cal M}}D_{i}^{A}D^{A i}$
with $D_{i}^{A}=\nabla_{i}-iA_{i}$ as covariant derivative.
Using the metric (3.2) and the given connection one-form,  
the Hamiltonian can be written as 
\begin{equation}
H= \frac{-\lambda}{{\cal M}}\{\frac{\partial^{2}}{\partial \theta^{2}} +
\gamma (\frac{1-2q}{\tan \gamma \theta}+\frac{2q}{\sin \gamma \theta}) 
\frac{\partial}{\partial \theta}+\frac{\gamma^{2}}{\sin^{2} \gamma \theta}
\frac{\partial^{2}}{\partial \Phi^{2}} +2iq \gamma^{2} (\frac{1}{\tan^{2}
\gamma \theta  \cos \gamma \theta}-\frac{1}{\sin^{2} \gamma \theta}) \frac{ 
\partial}{\partial \Phi}+q \gamma^{2} \},
\end{equation}
with the spectrum given by  
\begin{equation}
E(n)=\frac{\lambda \gamma^{2}}{{\cal M}}[n(n+1)-q(2n+1)],
\end{equation}
where in absence of angular deficit $n$ is a non-negative integer \cite{Jafar}.
Writting the eigenstates of the Hamitonian (3.4) in the following form 
$$
\Psi_{n,m}(\theta, \Phi)=(e^{-i\Phi}\frac{\tan \frac{\gamma \theta}{2}}{\sin \gamma \theta})^{-q}
e^{i\frac{m}{1-GM}\Phi}(1-\cos \gamma \theta)^{\frac{1}{2}|\frac{m}{1-GM}+q|}
(1+\cos \gamma \theta)^{\frac{1}{2}|\frac{m}{1-GM}-q|}F(\cos \gamma \theta)
$$
and choosing the change of variable $z=\frac{1-\cos \gamma \theta}{2}$,
the eigenvalue equation turns in to hypergeometric differential equation 
\begin{eqnarray}
z(1-z)\frac{d^2}{dz^2}F+[1+|\frac{m}{1-GM}+q|-(|\frac{m}{1-GM}-q|+
|\frac{m}{1-GM}+q|+2)z]\frac{d}{dz}F  \nonumber \\
+[n-q-\frac{1}{2}(|\frac{m}{1-GM}-q|+|\frac{m}{1-GM}+q|)]
[n-q+\frac{1}{2}(|\frac{m}{1-GM}-q|+|\frac{m}{1-GM}+q|)+1]F=0. \nonumber                    
\end{eqnarray}
The wave equations for relativistic and non-relativistic quantum states are
\renewcommand{\theequation}{\thesection.\arabic{equation}{a}}
\begin{equation}
H\Psi_{nonrel}(\theta,\Phi)=E(n)\Psi_{nonrel}(\theta,\Phi)   
\end{equation}
\vspace{-3mm}
\renewcommand{\theequation}{\thesection.\arabic{equation}{b}}
\setcounter{equation}{5}
\begin{equation}
H\Psi_{rel}(\theta,\Phi)=\frac{E_{rel}^{2}-{\cal M}^{2}}{2{\cal M}} 
\Psi_{rel}(\theta,\Phi). 
\end{equation}
For $\gamma=0$ we define $n \gamma=k$, which for $n$ very large $k$ is an
arbitrarily finite constant, and we get $E(n) \rightarrow \frac{k^{2}}{2{\cal M}}$. 
Then, with assumption $E_{rel}>{\cal M}$,
solutions of  Eqs. (3.6) are  
\renewcommand{\theequation}{\thesection.\arabic{equation}{a}}
\begin{equation}
\Psi_{_{nonrel}}(\theta,\Phi)=e^{i\frac{m}{1-GM}\Phi}J_{\frac{|m|}{1-GM}}(k\theta)
\end{equation}
\vspace{-3mm}
\renewcommand{\theequation}{\thesection.\arabic{equation}{b}}
\setcounter{equation}{6}
\begin{equation}
\Psi_{_{rel}}(\theta,\Phi)=e^{i\frac{m}{1-GM}\Phi}J_{\frac{|m|}{1-GM}}(\frac{1}{2\lambda}\sqrt{E_{rel}^{2}-{\cal M}^{2}} \theta).
\end{equation}
It is straitforward to see that by defining $\gamma=i\delta$ and taking the zero
limit of $\delta$ as $n\delta=k$ together with $E_{rel}<{\cal M}$, we get the
solutions of Eqs. (3.6) in terms of 
modified Bessel functions 
\renewcommand{\theequation}{\thesection.\arabic{equation}{a}}
\begin{equation}
\Psi_{_{nonrel}}(\theta,\Phi)=e^{i\frac{m}{1-GM}\Phi}K_{\frac{|m|}{1-GM}}(k\theta) 
\end{equation}
\vspace{-3mm}
\renewcommand{\theequation}{\thesection.\arabic{equation}{b}}
\setcounter{equation}{7}
\begin{equation}
\Psi_{_{rel}}(\theta,\Phi)=e^{i\frac{m}{1-GM}\Phi}K_{\frac{|m|}{1-GM}}(\frac{1}{2\lambda}\sqrt{{\cal M}^{2}-E_{rel}^{2}} \theta).   
\end{equation}
Therefore, in the presence of a heavy particle $M$ which is located at origin
we get scattering and bound states of a point particle with mass ${\cal M}$,
similar to Eqs. (3.7) and (3.8).
These results are in agreement with reference \cite{Krzysztof}.

So far the domain of $\Phi$ was $[0,2 \pi (1-GM)]$, but, from now on, for simplicity, we ignore  
the presence of this angular deficit in the rest of the article.
In general, in order to obtain the  eigen-spectrum algebraically, we introduce
generators of $gl(2,c)$ Lie algebra as 
\renewcommand{\theequation}{\thesection.\arabic{equation}}
\begin{eqnarray}
\nonumber
&&[L_{+}, L_{-}]=2\gamma^{2}L_{3}- 2q\gamma^{2}I \\ 
\nonumber
&&[L_{3}, L_{\pm}]=\pm L_{\pm}  \\
&&[{\bf L}, I]=0.   
\end{eqnarray}
Note that the algebra (3.9) becomes $iso(2) \oplus u(1)$ algebra for $\gamma=0$, 
$u(2)$ Lie algebra for $\gamma=1$, and $u(1,1)$ 
Lie algebra for $\gamma=i$.
The raising $L_{_{+}}$ and lowering $L_{_{-}}$ operators
have the following coordinate representations \cite{Jafar}
\begin{eqnarray}
L_{+}&=&e^{i\phi}(\frac{\partial}{\partial \theta}+\frac{i \gamma}{
\tan \gamma \theta}\frac{\partial}{\partial \phi})           
\hspace{54mm} L_{3}=\frac{1}{i}\frac{\partial}{\partial \phi} \nonumber \\
L_{-}&=&e^{-i\phi}(-\frac{\partial}{\partial \theta}+\frac{i\gamma}{
\tan \gamma \theta}\frac{\partial}{\partial \phi}+2q\frac{\gamma}{\tan
\gamma \theta}-2q\frac{\gamma}{\sin \gamma \theta})    
\hspace{10mm} I=1.
\end{eqnarray}
It is easy to show that the Hamiltonian (3.4) without angular deficit is
the Casimir operator of $gl(2,c)$ Lie algebra, that is 
\begin{eqnarray}
H=\frac{2\lambda}{M}[\frac{1}{4}L_{_{+}}L_{_{-}}+\frac{1}{4}
L_{_{-}}L_{_{+}}+\frac{1}{2}\gamma^{2}L_{_{3}}^{2}-q\gamma^{2}L_{_{3}}].
\end{eqnarray}
For $\gamma \neq 0$, its highest weight can be obtained  as follows
$$
\Psi_{n,n}(\theta,\phi)=e^{in\phi}(\frac{1}{\gamma} 
\sin \frac{\gamma \theta}{2})^{n}(\cos \frac{\gamma \theta}{2})^{n}, 
$$
where $n$ is a nonnegative integer of integration constant. The other states can be obtained by applying the lowering operator
$L_{-}$ over $\Psi_{n,n}(\theta,\phi)$ repeatedly. That is,   
\begin{eqnarray}
\Psi_{_{nonrel}}(\theta,\phi)&=&\Psi_{n,m}(\theta,\phi)=(L_{-})^{n-m}\Psi_{n,n}(\theta,\phi)  \nonumber  \\ 
&=&(2q-2n)_{n-m}e^{im \phi}(\frac{\sin \frac{\gamma \theta}{2}}{\gamma})
^{m} (\cos \frac{\gamma \theta}{2})^{2n-m} F(m-n, 2q-n, 2q-2n,
\frac{1}{\cos^{2}\frac{\gamma \theta}{2}}). \nonumber 
\end{eqnarray}
Hence, using the usual 
properties of hypergeometric function \cite{Vilenkin},
up to a constant coefficient $(-1)^{n-2q}(2q-2n)_{_{n-m}}$, 
in the limit $n \rightarrow \infty$ and $\gamma \rightarrow 0$
such that $n\gamma=k=$finite, the solution of algebraic method can
be written as 
\begin{eqnarray}
\nonumber
\Psi_{_{m}}(\theta,\phi)&=&\lim_{n \rightarrow \infty}k^{-m}e^{im\phi}
\cos^{4q}(\frac{k\theta}{2n})n^{m}\tan^{m}(\frac{k\theta}{2n})
F(2q-n,n+1,m+1,\sin^{2}(\frac{k\theta}{2n}))  \\
&=& k^{-m}e^{im\phi}\Gamma(m+1)J_{m}(k\theta), \nonumber
\end{eqnarray}
which is the same as special case $M=0$ in Eq. (3.7a). It is obvious that with 
the definition $\gamma=i \delta$, we will again get the solution (3.8a)
without angular deficit in the zero limit of $\delta$.

For the left and right
invariant generators of $SL(2,c)$ group manifold with $sl(2,c)$ Lie algebra
\begin{eqnarray}
\nonumber
&&J_{_{\pm}}^{(L)}=e^{\pm i\phi}
(\pm \frac{\partial}{\partial \theta}+i\frac{\gamma}{\tan \gamma \theta} 
\frac{\partial}{\partial \phi}-i\frac{\gamma}{\sin \gamma \theta} 
\frac{\partial}{\partial \psi})  \\
&&J_{_{3}}^{(L)}=-i\frac{\partial}{\partial \phi},
\end{eqnarray}     
\begin{eqnarray}
\nonumber
&&J_{_{\pm}}^{(R)}=e^{\pm i\psi}
(\pm \frac{\partial}{\partial \theta}-i\frac{\gamma}{\sin \gamma \theta}
\frac{\partial}{\partial \phi}+i\frac{\gamma}{\tan \gamma \theta}
\frac{\partial}{\partial \psi}) \\
&&J_{_{3}}^{(R)}=-i\frac{\partial}{\partial \psi},
\end{eqnarray}
where $\theta$, $\phi$ and $\psi$ are complex variables at present,
the Casimir operators are equal with each other, i.e.
$H_{_{sl(2,c)}}=H^{(L)}=H^{(R)}$, such that
\begin{eqnarray}
H^{(L,R)}=
\frac{1}{4}J_{_{+}}^{(L,R)}J_{_{-}}^{(L,R)}+\frac{1}{4}J_{_{-}}^{(L,R)}
J_{_{+}}^{(L,R)}+\frac{1}{2}\gamma^{2}J_{_{3}}^{{(L,R)}^2}.
\end{eqnarray}
Since $so(4,c)=sl(2,c) \oplus sl(2,c)$, therefore the Casimir operator of
$so(4,c)$, can be given in terms of the $sl(2,c)$ Casimir operators as
$H_{_{so(4,c)}}=H^{(L)}+H^{(R)}$,
but note that, choosing the coordinates of left and right generators the same
will be equivalent to restricting the space of $so(4,c)$ Lie algebra
to one of the subspaces $sl(2,c)$.
Now we restrict ourselves to real values of $\theta$, $\phi$ and $\psi$
variables such that $0 \leq \phi < 2\pi$ and $0 \leq \psi < 4\pi$
for all values of $\gamma$, while we choose $0 \leq \theta < \pi$ for $\gamma=1$
and $0 \leq \theta < \infty$ for other cases.
Due to this restriction, the $sl(2,c)$ algebra reduces to its different real 
forms as follows:
in the case of $\gamma=1$ the $sl(2,c)$ Lie algebra reduces
to $su(2)$, while for $\gamma=i$ it reduces to $su(1,1)$ Lie algebra and
finally for $\gamma=0$ it reduces to $iso(2)$ Lie algebra \cite{Gilmore}. We should remind that
for $\gamma=1$, $\gamma=i$ and $\gamma=0$ the direct sum of left and right invariant
generators becomes $so(4)=su(2) \oplus su(2)$, $so(2,2)=su(1,1) \oplus su(1,1)$
and $iso(2) \oplus iso(2)$ Lie algebra respectively, which are different real forms of $so(4,c)$.

If we define $l:=n-q$, the eigenvalue equation for  
Casimir operator $H_{_{sl(2,c)}}$  will be \cite{Vilenkin}
\begin{equation}
H_{_{sl(2,c)}} \Psi_{_{l,m,q}}(\theta,\phi,\psi)=\frac{1}{2} \gamma^{2}
l(l+1) \Psi_{_{l,m,q}}(\theta,\phi,\psi)
\end{equation}
with
$\Psi_{_{l,m,q}}(\theta,\phi,\psi)=e^{im \phi-iq \psi} P_{_{m,-q}}^{l}(\cos
\gamma \theta).$
After transfer the $\psi$ dependence factor $e^{-iq \psi}$ to the left
hand side of equation (3.15), the reduced Casimir operator with degeneracy group $SL(2,c)$, is 
\begin{eqnarray}
H_{_{sl(2,c)}}(q)=\frac{1}{4}J_{_{+}}^{(L)}(q)J_{_{-}}^{(L)}(q)+\frac{1}{4}
J_{_{-}}^{(L)}(q)J_{_{+}}^{(L)}(q)+\frac{1}{2}\gamma^{2}J_{_{3}}^{{(L)}^2}(q),
\end{eqnarray}
where
\begin{eqnarray}
&& J_{_{\pm}}^{(L)}(q)=e^{\pm i\phi}(\pm \frac{\partial}{\partial \theta}+\frac{i
\gamma}{\tan \gamma \theta} \frac{\partial}{\partial \phi}- \frac{q\gamma}{
\sin \gamma \theta})  \nonumber \\
&& J_{_{3}}^{(L)}(q)=-i\frac{\partial}{\partial \phi}. \nonumber
\end{eqnarray}
The above reduced $sl(2,c)$ generators are related to (3.10) 
generators of $gl(2,c)$  through a similarity transformation together with
a change of basis,
such that, one can obtain the eigenstates of $gl(2,c)$ Casimir operator.
Thus, the eigenvalue equation (3.15) reduces to
\begin{equation}
H_{_{sl(2,c)}}(q)\Psi_{_{l,m,q}}(\theta,\phi)=\frac{1}{2}\gamma^{2}l(l+1)
\Psi_{_{l,m,q}}(\theta,\phi),
\end{equation}
with $\Psi_{_{l,m,q}}(\theta,\phi)=e^{im\phi}P_{_{m,-q}}^{l}(\cos\gamma\theta)$.
From there we can introduce the eigenfunction
\begin{eqnarray}
\Psi_{n,m}(\theta,\phi)=
(2\frac{\tan \frac{\gamma \theta}{2}}{\sin \gamma \theta}e^{-i\phi})^{-q}
e^{im\phi}P_{_{m,-q}}^{l}(\cos\gamma\theta), \nonumber
\end{eqnarray}
for Hamiltonian (3.11) with the same eigenvalue (3.15).
Then, we see that the Dirac's quantization of magnetic charge 
follows very naturally from the $SL(2,c)$ representation,
that is integrity of $q$.
Also, the $so(4,c)$ Lie algebra is the
dynamical group of the quantum dynamics of a point mass particle in the 
presence of a constant magnetic 
field over two dimensional surface with global constant curvature with $gl(2,c)$ 
sub-algebra as its degeneracy group.

Transferring the function $e^{-iq\psi}$ to the left of the Casimir operator
$H_{_{sl(2,c)}}$
in the eigenvalue equation (3.15),  
the Casimir operator $H_{_{sl(2,c)}}(q)$
can be written as
\renewcommand{\theequation}{\thesection.\arabic{equation}{a}}
\begin{equation}
H_{_{sl(2,c)}}(q)=\frac{1}{2}J_{_{+}}^{(R)}(q+1)J_{_{-}}^{(R)}(q)+\frac{1}{
2}q(q+1)\gamma^{2}    
\end{equation}
\vspace{-7mm}
\renewcommand{\theequation}{\thesection.\arabic{equation}{b}}
\setcounter{equation}{17}
\begin{equation}
=\frac{1}{2}J_{_{-}}^{(R)}(q-1)J_{_{+}}^{(R)}(q)+\frac{1}{2}q(q-1)\gamma^{2}
\end{equation}
with
\renewcommand{\theequation}{\thesection.\arabic{equation}}
\begin{eqnarray}
J_{_{\pm}}^{(R)}(q)=\pm \frac{\partial}{\partial \theta}-i\frac{
\gamma}{\sin \gamma \theta} \frac{\partial}{\partial \phi}+\frac{q\gamma}{
\tan \gamma \theta}. \nonumber
\end{eqnarray}
Using the Eqs. (3.18) in the eigenvalue equation (3.17)  
we obtain
\begin{eqnarray}
&&J_{_{+}}^{(R)}(q)J_{_{-}}^{(R)}(q-1)\Psi_{_{l,m,q-1}}(\theta,\phi)=E_{_{q}}\Psi_{_{l,m,q-1}}(\theta,\phi)   \nonumber   \\
&&J_{_{-}}^{(R)}(q-1)J_{_{+}}^{(R)}(q)\Psi_{_{l,m,q}}(\theta,\phi)=E_{_{q}}\Psi_{_{l,m,q}}(\theta,\phi),        \nonumber
\end{eqnarray}
where $E_{_{q}}$ is defined as 
\begin{eqnarray}
E_{_{q}}=\gamma^2[l(l+1)-q(q-1)]. \nonumber
\end{eqnarray}
The restriction of $so(4,c)$ Lie algebra to $sl(2,c)$, together with
$H_{_{sl(2,c)}}$ reduction, lead to eigenvalue equation (3.17) which can
be obtained from Hamiltonian (3.11) via similarity transformation. The
Hamiltonian $H_{_{sl(2,c)}}(q)$, on the one hand, is quadratic Casimir operator of $sl(2,c)$
Lie algebra with  $J_{_{\pm}}^{(L)}(q)$ and  $J_{_{3}}^{(L)}(q)$ generators
which, as a result, possesses $gl(2,c)$ degeneracy symmetry. On the other hand,
writting the Hamiltonian in terms of $J_{_{\pm}}^{(R)}(q)$ will possess the shape invariance
symmetry. Both properties are related to the realization of
para-supersymmetry of arbitrary order \cite{Jafar,Jaf}.

\vspace{7mm}

\begin{center}
{\bf {\large  ACKNOWLEDGEMENT}}
\end{center}
We wish to thank  Dr. S.K.A. Seyed Yagoobi for carefully reading the 
article and for his constructive comments.

\end{document}